\documentclass[12pt,english,iopart]{article}
\usepackage[T1]{fontenc}
\usepackage[latin1]{inputenc}
\usepackage{graphicx}

\makeatletter

\newcommand{\lyxaddress}[1]{
\par {\raggedright #1
\vspace{1.4em}
\noindent\par}
}

\usepackage{times}

\usepackage{geometry}

\usepackage{setspace}

\makeatletter

\date{}
\usepackage{cite}

\makeatother

\usepackage{babel}
\makeatother
\begin{document}

\title{Robust nanopatterning by laser-induced dewetting of metal nanofilms}

\author{$^{1,2}$Christopher Favazza, $^{1,2}$Ramki Kalyanaraman%
\thanks{Corresponding author, ramkik@wuphys.wustl.edu%
} , $^{2,3}$Radhakrishna Sureshkumar%
\thanks{E-mail: suresh@che.wustl.edu%
}}

\maketitle

\lyxaddress{\begin{center}$^{1}$Department of Physics, Washington University
in St. Louis, MO 63130\\
 $^{2}$Center for Materials Innovation, Washington University in
St. Louis, MO 63130\\
 $^{3}$Department of Chemical Engineering, Washington University
in St. Louis, MO 63130\par\end{center}}

\begin{abstract}
We have observed nanopattern formation with robust and controllable
spatial ordering by laser-induced dewetting in nanoscopic metal films.
Pattern evolution in Co film of thickness $1\leq h\leq8\, nm$ on
$SiO_{2}$ was achieved under multiple pulse irradiation using a 9
ns pulse laser. Dewetting leads to the formation of cellular patterns
which evolve into polygons that eventually break up into  nanoparticles
with monomodal size distribution and short range ordering in nearest-neighbour
spacing $R$. Spatial ordering was attributed to a hydrodynamic thin
film instability and resulted in a predictable variation of $R$ and
particle diameter $D$ with  $h$. The length scales $R$ and $D$
were found to be independent of the laser energy. These results suggest
that spatially ordered metal nanoparticles can be robustly assembled
by laser-induced dewetting.\pagebreak
\end{abstract}

\section{Introduction}

Ordered metal nanoparticle arrays have become an active area of research
due to their potential applications in nonlinear optics and nanophotonics
below the diffraction limit \cite{quinten98,lamprecht00,maier03}.
 However, practical realization of such applications require that
cost-effective and reliable processes be  developed for the manufacturing
of ordered metal nanoarrays. Physical phenomena that  lead to the
formation of patterns with characteristic length scales and predictable
time scales could be exploited in designing such nanomanufacturing
processes. For instance, pattern formation following ion irradiation
has been known to modify surface topography \cite{erlebacher00,facsko01,valbusa03,teichert04,buatier05}.
Rippling of surfaces under ion irradiation appears to be primarily
due to an instability resulting from the competition between ion erosion
and smoothening due to surface diffusion with the length and time
scales determined by ion flux, surface temperature and surface diffusion
parameters \cite{Sigmund65,harper88,Chason94,erlebacher99}\textbf{.}
 Another route to creating nanopatterns could be through instabilities
of a thin fluid film leading to spatial patterns with length and time
scales that depend on the thermophysical material properties such
as interfacial tension, contact angle with the substrate, fluid viscosity
and, for ultrathin films, long range dispersion forces such as the
van der Waal's interaction.  An example especially pertinent to thin
film pattern formation is the hydrodynamic dewetting instability such
as  spinodal dewetting which occurs when attractive intermolecular
forces exceed the stabilizing effect of interfacial tension \cite{vrij66,sharma86,sharma93,Seemann01}.
Under such conditions, spontaneous film thickness fluctuations  could
be amplified and this will result in the breakup of the film leading
eventually to the formation of drops/particles with well defined spatial
order \cite{Wyart90}. Dewetting dynamics leading to particles have
been studied in detail in polymer films that are in the liquid state
close to room temperature \cite{reiter92,thiele98,stange97,thiele01}.
However, ordered nanoparticle formation via dewetting in metal films
has remained relatively unexplored , primarily due to the high temperatures
required to melt the metal and observe dewetting within practical
time scales. Earlier work has shown that ion-irradiation of ultrathin
Pt films on $SiO_{2}$ substrates results in dewetting patterns with
spatial order \cite{Averback00,averback01}. More recently our group
has shown that multiple $ns$ pulsed laser melting of nanoscopic Co
films can lead to short- and long-range spatial order \cite{Favazza06a,favazza06b,favazza06c}.
However, the the relationship of particle spacing $R$ and diameter
$D$ with film thickness $h$ was not determined. Furthermore, it
is not clear whether thin film hydrodynamic models could help predict
such relationships at nanometer length scales. Another unanswered
question is whether dewetting under multiple cycles of phase change,
i.e. melting and resolidification, would lead to robust final states.
In this article, we address the above questions quantitatively by
analyzing the results of a series of systematic experiments and interpreting
them with the aid of thin film hydrodynamic theories. Such comprehensive
understanding could help in knowledge-based design of nanomanufacturing
facilities for ordered metal nanoarrays. 

Specifically, we investigate the evolution of dewetting patterns in
nanoscopic $Co$ metal films  on $SiO_{2}$ in vacuum under multiple
pulse irradiation with a laser of pulse length $\tau_{p}=9\, ns$
and wavelength $266\, nm$. Co nanoparticles are useful towards nanoscale
magnetic applications and in catalysis. Dewetting was studied as a
function of film thickness $h$, laser energy density $E$ and laser
irradiation time expressed in terms of the number of pulses $n$.
We discovered that pattern formation evolved from cellular structures
composed of polygons at early stages, which coalesced and eventually
changed into nanoparticles at later times. Every observed pattern
was ordered with a characteristic length scale while the nanoparticles
had a well-defined nearest-neighbour ($NN$) spacing $R$ and a monomodal
size distribution with an average diameter $D$. The observed variations
of $R$ and $D$ as function of $h$ were in agreement with the linear
theory of spinodal dewetting. We also found that $R$ was \emph{independent}
of $n$ and $E$, provided net $Co$ evaporation was small. Based
on these observations we infer that once the pulsed laser melting
initiated dewetting, subsequent pulses evolved this instability while
maintaining its spatial characteristics, thereby leading to a predictable
final state\emph{.} This offered strong evidence that the dewetting
dynamics is extremely robust. \emph{To our knowledge this is the first
experimental evidence for: (i) formation of metal nanoparticles with
spatial order via a hydrodynamic dewetting instability; (ii) evidence
for the linear spinodal theory under multiple instances of phase change;
and (iii) evidence for robustness of the dewetting process to experimental
conditions.}

\section{Experimental Details}

$Co$ films with thickness $\sim1$ to $8$ nm were deposited at rates
varying between $0.5\, to\,2\: nm/min$ onto commercial optical quality
$SiO_{2}/Si$ wafers consisting of $400\, nm$ thick thermally grown
oxide layer on polished $Si(100)$ wafers under high vacuum $(2\times10^{-8}\, Torr)$
by e-beam evaporation at room temperature \cite{favazza06c,Favazza06a}.
For every film thickness we measured the surface roughness and average
grain size. Atomic force microscopy (AFM) of the surface roughness
established an upper limit of $0.1\pm.03\, nm$ average rms roughness
for the entire thickness range while transmission electron microscopy
gave a average grain size of $\sim20\, nm$. No systematic change
in roughness or particle size was detected with increasing film thickness.
Further, no spatially ordered defects or other heterogeneities were
observed on the native substrate surface. For the film thickness range
investigated the laser absorption at $266\, nm$ wavelength was uniform
within the Co film (absorption depth $\sim11\, nm$).  Therefore vertical
temperature gradients could be neglected and the entire film temperature
may be modeled as a uniform, time varying function that depends on
the laser power (source term) and conduction of heat to the substrate
(sink term) \cite{Trice06a}. The film thickness was measured by concentration
measurements of Co via energy dispersive X-ray spectrometry (EDS)
in a scanning electron microscope (SEM). The concentration was converted
into an equivalent thickness value by using calibration based on  step-height
and Rutherford backscattering measurements of the film thickness.
The films were irradiated immediately after deposition in vacuum at
normal incidence by an unfocused laser beam of area $3\times3\, mm^{2}$
at a repetition rate of 50 Hz. Under these conditions, the film surface
was never exposed to air. Hence  no oxide layer was expected to form
on the film surface. Pattern formation was studied for energy densities
$E$ between $40\leq E\leq150\, mJ/cm^{2}$ and for $n$ between 10
and 10,500 pulses. For each thickness the range of $E$ was chosen
to be above the melt threshold. This was determined by a visible roughening
of the metal film surface, as detected under high-resolution SEM within
the longest time scale of the experiment (i.e. after $10,500$ laser
pulses) \cite{matthias94}. Analytical estimates based on a 1-D heat
transport model as well as finite element simulations showed that
the lifetime $\tau_{m}$ of the Co liquid ranged from $\sim2\leq\tau_{m}\leq12\, ns$
in the entire $h$ range studied \cite{Trice06a}. The same model
predicted cooling rates of the order of $10^{10}K/s$ and a total
heating plus cooling time per pulse of $\sim100\, ns$, which was
much smaller than the spacing between pulses of $20\, ms$. Hence,
during each pulse, the film melts and in the molten state, dewetting
process occurs. Upon the cessation of the pulse, the film solidifies
within a time scale much smaller than the interval between subsequent
pulses. The maximum temperature of the molten liquid above the Co
melting point $T_{m}=1768\, K$ was estimated to be $\sim500\, K$.
Under these conditions, the maximum evaporation observed after the
longest $n$ at the highest $E$ was $\leq15\%$.

\section{Results }

Fig. \ref{cap:Patternevolution} shows the dewetting morphology observed
for irradiation at $E=120\, mJ/cm^{2}$ of a $3.8\, nm$ thick Co
film for increasing number of laser pulses $n$. This evolution sequence
is typical of the ordered morphology observed in the thickness range
of $1\leq h\leq8$ nm. After a few pulses, the dewetting morphology
is characterized by a cellular network of polygons (Fig. \ref{cap:Patternevolution}A).
As $n$ is increased  the metal/substrate contact lines recede to
the edge of the holes resulting in a network of coalescing polygonal
features (Fig. \ref{cap:Patternevolution}B). Continued irradiation
results in the formation of nanoparticles primarily at the junctions
of the polygons, as evident from the location of the particles  in
Fig. \ref{cap:Patternevolution}B and in C and D. \emph{At every observed
stage, a characteristic length scale $L$ was present, as evidenced
by the annular form of the power spectrum of the spatial correlations
in the intensity variation within each pattern. The power spectrum
is shown alongside the corresponding images in Fig. \ref{cap:Patternevolution}A
to D.} For patterns consisting of polygons, $L$ represents the average
distance between the center to the vertices, while in the case of
the nanoparticles it represents the average $NN$ spacing $R$.

A nonlinear peak fitting algorithm (Levenberg-Marquardt) was used
to locate Gaussian shaped peaks in the radial distribution function
$g(k)$ versus the wavenumber $k\equiv1/L$ for each pattern (Fig.
\ref{cap:Timedependence}A). The variation of $L$ with $n$ for various
$E$ values is shown in Fig. \ref{cap:Timedependence}B (\emph{the
lines are only guides for the eye}). The initial rapid increase in
$L$ corresponds to the increase in the average size of the polygons.
The decrease corresponds to the breakup of polygons into nanoparticles,
with the long time value representing $R$. The size distribution
of nanoparticles in the final state was observed to be monomodal,
as shown in Fig. \ref{cap:Timedependence}C . Also, as seen from  Fig.
\ref{cap:Timedependence}B, the value of $R$ remains unchanged over
long irradiation times and is independent of laser energy $E$. However,
increasing $E$ increases the rate at which the patterns evolve, as
evident from Fig. \ref{cap:Timedependence}B and the figure in the
inset. At the lowest energy of $110\, mJ/cm^{2}$ the time to reach
$R$ ( $n>3000$) was observed to be much larger than the $130\, mJ/cm^{2}$
case ( $n\leq3000$). Since the temporal scale used here is coarse,
i.e., it is expressed in number of pulses,   the precise location
of the peak value of $L$ was not detectable for every energy value.

\section{Discussion}

It is well accepted that dewetting morphology can progress via three
pathways \cite{degennes03}: (i) Homogeneous nucleation and growth,
where holes appear spontaneously at random locations and times on
the surface.  Because of the inherent randomness, no characteristic
length is present in this type of dewetting \cite{stange97}; (ii)
 Heterogeneous nucleation and growth due to defects, impurities or
other experimentally imposed heterogeneities. In this type of dewetting,
a characteristic length scale could appear at the early stages of
dewetting due to ordered nucleation sites. For instance, in ion-irradiation
induced dewetting, the average molten zone of an ion imposes a characteristic
length scale in dewetting \cite{averback01}; and (iii)  Thin film
hydrodynamic (T.F.H.) instabilities  such as the one associated with
the dewetting of spinodally unstable systems. The resulting patterns
are  characterized by a well-defined length scale in the hole spacing
and/or size \cite{thiele98}. Since we have ruled out spatially ordered
heterogeneities on the surface and in the film microstructure, the
evidence for spatial ordering in Fig. \ref{cap:Patternevolution}
strongly suggests spinodal dewetting.

We examined whether T.F.H. models developed based on lubrication analysis
 could help explain the time and length scales of the observed dewetting
patterns. Towards this end, we adopted the T.F.H. equations employed
 by Vrij \cite{vrij66,vrij68} and Sharma and Ruckenstein \cite{sharma86}
to study spinodal dewetting in ultrathin polymeric films. Lubrication
theory provides a relationship between the mass flux (which can be
related to temporal variations in film height) and the effective pressure
gradient, which for a thin film could result from perturbations that
produce interface curvature, and disjoining pressure due to  long-range
intermolecular forces such as van der Waals interactions. For a thin
film resting on a substrate contribution to the disjoining pressure
can be expressed as  $A/6\pi h'^{3}$ where \emph{A} is the Hamaker
constant and $h'$ is the instantaneous film thickness. This approach
leads to the dynamical equation describing the film height as a function
of time as:\begin{equation}
3\eta\frac{\partial h'}{\partial t}=-\bigtriangledown.\left(\gamma h'^{3}\bigtriangledown.\bigtriangledown^{2}h'+\frac{A*}{h'}\bigtriangledown h'\right)\label{eq:DynEq}\end{equation}

where $\eta$ is the viscosity and $\gamma$ is the interfacial energy
of the film-vacuum interface and $A*\equiv A/2\pi$. If infinitesimally
small perturbations $\hat{h}$ are imposed on an initially flat film
of uniform thickness \emph{h}, i.e. \emph{}$h'=h+\hat{h}$, the growth/decay
rate of such perturbations can be assessed by using linear stability
analysis in which one keeps only terms which are linear in $\hat{h}$:
\begin{equation}
3\eta\frac{\partial\hat{h}}{\partial t}=-\nabla.\left(\gamma h^{3}\nabla.\nabla^{2}\hat{h}\,+\frac{A*}{h}\nabla\hat{h}\right).\label{eq:hat}\end{equation}

Substitution of the normal mode form $\hat{h}=\epsilon\exp(ik_{x}x+ik_{y}y+\sigma t)$,
where \emph{x} and \emph{y} denote the orthogonal surface coordinates
and $\sigma$ represents the growth or decay rate of perturbations
with the characteristic wavevector $\mathbf{k}\equiv(k_{x},\, k_{y})$,
into eq. \ref{eq:hat} gives the following dispersion relationship
($k^{2}\equiv\mathbf{k}.\mathbf{k)}$:\begin{equation}
3\eta\sigma=-\gamma h^{3}k^{4}+\frac{A*}{h}k^{2}.\label{eq:dispersion}\end{equation}
 It is evident from eq. \ref{eq:dispersion} that $\sigma>0$ for
perturbations with wavevectors with $k^{2}<\frac{A*}{\gamma h^{4}}$
implying that such perturbations will be amplified as time progresses.
The fastest growing mode $k_{m}$ can be obtained by letting $\frac{\partial\sigma}{\partial k}=0.$
This gives $k_{m}^{2}=\frac{A*}{2\gamma h^{4}}$. Hence, the   characteristic
wavelength $\Lambda\equiv2\pi/k_{m}$  is given by:\begin{equation}
\Lambda=\sqrt{\frac{16\pi^{3}\gamma}{A}}\,\, h^{2}.\label{eq:Lambda}\end{equation}
The time scale $\tau_{D}\equiv2\pi/\sigma$ associated with the growth
of perturbations with wavelength $\Lambda$ can be evaluated by letting
$k=k_{m}$ in eq. \ref{eq:dispersion} as:\begin{equation}
\tau_{D}=\frac{96\pi^{3}\gamma\eta h^{5}}{A^{2}}.\label{eq:TauD}\end{equation}
 A previous investigation by Bischof et al.  using single shot ns
pulsed laser melting of a multilayer metal structure has shown that
is possible to capture snapshots of a dewetting instability provided
the characteristic time scale of the dewetting process is much larger
than the liquid melt time $\tau_{m}$ \cite{bischof96,bischof96b}.
Based on our analytical model  the liquid lifetimes were estimated
to be $2\leq\tau_{L}\leq12\, ns$ for the range of experimental parameters
reported here \cite{Trice06a}.  Further, using eq. \ref{eq:TauD}
we estimated  the dewetting time scale at the melting point $T_{melt}$
of Co using $\gamma^{Co}(T_{m})\,=1.88\, J/m^{2}$, viscosity $\eta^{Co}(T_{m})=4.46\times10^{-3}\, Pa-s$,
and a typical range of the \emph{Hamaker} constant $A$ for metals
from $10^{-20}\leq A\leq10^{-18}\, J$. We found that the lower limit
is $\tau_{D}\sim25\, ns$ for a $1\, nm$ thick film and this time
increases rapidly with increasing $h$. Therefore, it is reasonable
to expect that pulsed laser melting used in this experiment could
capture snapshots of spinodal dewetting. We further suggest that once
the dewetting instability is initiated, continued laser pulsing fosters
the instability  eventually leading to a robust final state whose
length scale, as suggested by eq. \ref{eq:Lambda}, should be independent
of the laser parameters. This effect is observed in Fig. \ref{cap:Timedependence}B,
where despite a change in laser $E$ and/or $n$, an identical $NN$
spacing $R$ is observed in the stable nanoparticle state.

To provide further quantitative evidence for spinodal dewetting, we
have investigated the experimental variation of the $NN$ particle
spacing $R$ with $h$. The origin of the space-filling polygonal
shapes can be explained by the presence of the preferred spinodal
wavenumber $k\sim\frac{1}{\Lambda}$, which is a scalar measure of
the wavevectors $k_{x}$ and $k_{y}$ in the two orthogonal directions
along the undisturbed planar film, defined as $k=\sqrt{k_{x}^{2}+k_{y}^{2}}$.
Since there are no external length scales in the $x$ and $y$ directions,
one could have various combinations of $k_{x}$ and $k_{y}$ leading
to the observed shapes \cite{Chandrasekhar81}. Therefore the result
of the linear spinodal theory (eq. \ref{eq:Lambda}) suggests that
the polygon characteristics such as  size, center-to-center spacings
and the vertex separations will be proportional to $h^{2}$. As we
have shown from Fig. \ref{cap:Patternevolution}, the observed nanoparticles
form predominantly at the polygon vertices. Therefore it is clear
that if the nanoparticle spacing is a result of spinodal dewetting,
then $R$ should be proportional to $\Lambda$ and also vary as $h^{2}$.
In Fig. \ref{cap:Timedependence}(D), the experimentally observed
variation of $R$ (solid circles) with $h$ is plotted. A power law
fit to the variation in $R$ yields $R=25.7h^{1.98\pm0.3}$ where
the exponent of $h$ is in excellent agreement with the prediction
of eq. \ref{eq:Lambda} of $2$. The $NN$ spacing varied from $\sim50$
to $1000\, nm$ for the films studied. From the prefactor of $25.7\, nm^{-1}$,
the Hamaker constant was estimated as $A=14.1\times10^{-19}J$ (assuming
the proportionality factor correlating $R$ to $\Lambda$ is 1). This
value is of the right sign and order of magnitude for metals in general
\cite{israelachvili92}. We also evaluated the average diameter $D$
of the nanoparticles as a function of $h$ (Fig. \ref{cap:Timedependence}D
open squares). $D$ varied from $\sim30$ to $250\, nm$ for the thickness
range studied here. Using a volume conservation argument $D$ should
vary with $h$ as $D\sim h^{5/3}$ \cite{reiter92}. The power-law
fit to our experimental data (Fig. \ref{cap:Timedependence}D) gave
$D=14.1h^{1.6\pm0.3}$ where the exponent is once again in good agreement
with the theoretical prediction. \emph{These results clearly support
a dewetting mechanism caused by a thin film hydrodynamic instability
and also provides evidence that the linear spinodal theory is applicable
 under experimental conditions in which multiple  cycles of film melting
and solidification are present.}

The break-up of the polygonal structures into droplets requires some
elaboration. Previous studies of break-up of cellular structures into
particles have attributed the process to a Rayleigh instability \cite{Rayleigh1879}
as a result of which the particle diameter will follow the trend $D\sim h^{3/2}$
\cite{reiter92,henley05}. However, our interpretation of the experimental
observations reported here is that particle formation occurs via progression
of the spinodal dewetting instability and through capillary effects.
The primary evidence that supports this interpretation is that the
nanoparticles predominantly form at the junctions of the polygons,
as compared to the classical string-like particle distribution resulting
from the Rayleigh instability of the polygon sides \cite{reiter92}.
Capillary effects resulting from the difference in curvature of the
spheroidal particles at the junctions and the polygon walls will also
act to drive the liquid towards the junction, a situation that is
relatively well understood from the classical problem of dragging
of a liquid film by a moving plate \cite{landau42}.

For the spinodal process it is possible to rationalize why $R$ is
invariant to laser energy $E$ despite the increase in the rate of
dewetting $\sigma\propto L/n$, as observed from Fig. \ref{cap:Timedependence}(B)
for early stages (small values of $n$). The increase in rate with
increasing $E$ is due to two effects: one is the increase in liquid
lifetime $\tau_{L}$ which will prolong the dewetting process within
a given number of pulses; the second is the increase in liquid temperature
$T_{L}$ above $T_{m}$which will increase the rate as:\[
\Delta\sigma(\Delta T)=-\sigma(T_{m})[\frac{1}{\gamma(T_{m})}\frac{d\gamma}{dT}+\frac{1}{\eta(T_{m})}\frac{d\eta}{dT}]\Delta T\]

where the Hamaker constant was treated as T-independent based on its
extremely weak T-dependence \cite{french00}. Since the the T-dependent
material parameters for Co of $d\gamma/dT=-0.5\times10^{-3}J/m^{2}-s$
; and $d\eta/dT=-6.3\times10^{-6}Pa-s/K$ both decrease with increasing
T, therefore, the rate \emph{increases with increasing T}, as qualitatively
observed from the inset of Fig. \ref{cap:Timedependence}(B). This
increase will be reinforced by that produced by the longer melt lifetime\emph{.}
A similar analysis shows that the dewetting length scales $L\propto\Lambda$
will decrease with increasing $\Delta T$ as:

\[
\Delta L(\Delta T)=\frac{L(T_{m})}{2\gamma(T_{m})}\frac{d\gamma}{dT}\Delta T=-2.66\times10^{-4}L(T_{m})\Delta T\]

This decrease in $L$ with increasing $T_{L}$ can be estimated to
be $\sim-13\%$ for $\Delta T=500\, K$, which is a relatively small
change and will be undetectable given the larger spread in the experimental
values of $R$ of $\pm20\%$ as measured from the $g(k)$. Therefore,
for conditions under which the net Co evaporation is small, the changes
in $R$ will be experimentally indistinguishable.

\section{Conclusion}

In conclusion, we studied pattern evolution in nanoscopic Co films
$(1\leq h\leq8\, nm)$ under irradiation by multiple $9\, ns$ laser
pulses. Laser melting initiates a dewetting instability and subsequent
irradiation fosters this dewetting to a stable state of spatially
ordered nanoparticles. The dewetting instability appears to be spinodal
in nature based on the observations of a characteristic length scale
at all stages of dewetting and the $\sim h^{2}$ behavior of the $NN$
spacing of the nanoparticles. \emph{To our knowledge this is the first
experimental evidence for: (i) formation of metal nanoparticles with
spatial order via spinodal dewetting; (ii) evidence for linear spinodal
theory under multiple cycles of phase change; and (iii) evidence for
robustness of the dewetting process with respect to experimental conditions.}
The short range order and particle size produced by the pattern selection
mechanisms can be manipulated by employing films with varying thickness.
This offers opportunities to develop cost-effective and easily controllable
nanomanufacturing processes for metal nanoarrays.

RK acknowledges support by the National Science Foundation through
grant \# DMI-0449258. The authors acknowledge invaluable comments
provided by Prof.'s Ken Kelton and Linday Greer during the preparation
of this manuscript.

\bibliographystyle{unsrt}

\pagebreak

\pagestyle{empty}

\section*{Figure captions}

\begin{figure}[t]

\begin{centering}\includegraphics[height=8in,keepaspectratio]{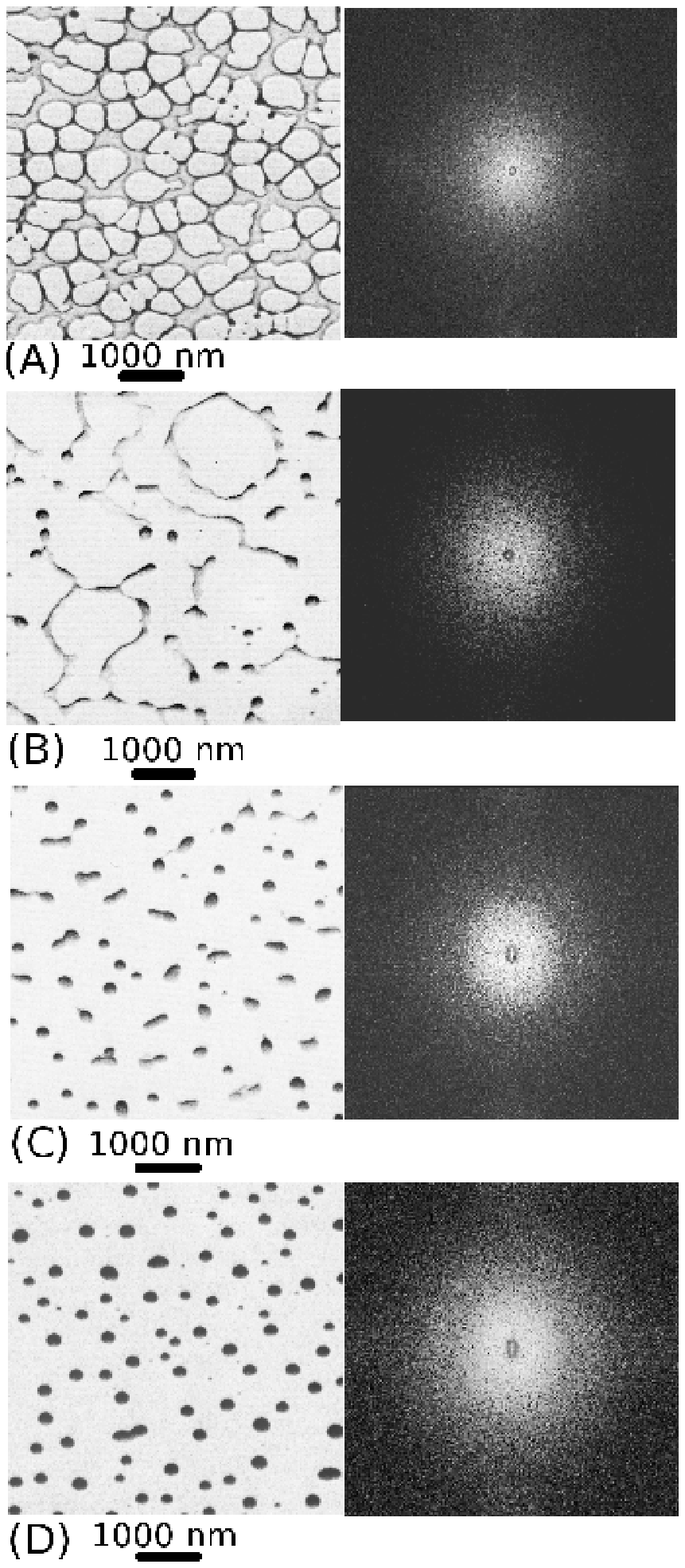}\par\end{centering}

\caption{\label{cap:Patternevolution}}
\end{figure}

\begin{itemize}
\item \textbf{Figure \ref{cap:Patternevolution}}: Scanning electron micrgraphs of dewetting patterns as a function of increasing number of laser pulses $n$ for
a $3.8\, nm$ thick $Co$ film at a laser energy density of $120\, mJ/cm^{2}$.
(A) Polygons after $n=10$; (B) Coalescing polygons with evidence
for particles at the junctions after $n=500$; (C) Nanoparticles with
some remaining evidence of the polygon edges after $n=3000$; and
(D) The final nanoparticle morphology after $n=10500$. The inset
of each figure shows the corresponding power spectra for each pattern.
 
\item \textbf{Figure \ref{cap:Timedependence}}:Evidence for the characteristic
length scales and their behavior for the dewetting process. (A) The
$g(k)$ (symbols) and fits (lines) for the patterns of Fig. \ref{cap:Patternevolution}.
(B) Dependence of $L$ on $n$ at various laser energies for the $3.8\, nm$
film. \emph{The dashed lines are only intended to be guides for the
eye}. The inset shows the change in L for various energies at early
stages, as measured after 10 pulses. (C) The monomodal size distribution
of nanoparticles in the final state for the $3.8\, nm$ film. (D)
Nearest-neighbour spacing $R$ (solid circles) and Diameter $D$ (open
squares) of the nanoparticles vs. film thickness $h$ and the least-squares
fit. \label{cap:Timedependence} 
\end{itemize}

\begin{figure}[t]

\begin{centering}\includegraphics[width=2in,keepaspectratio]{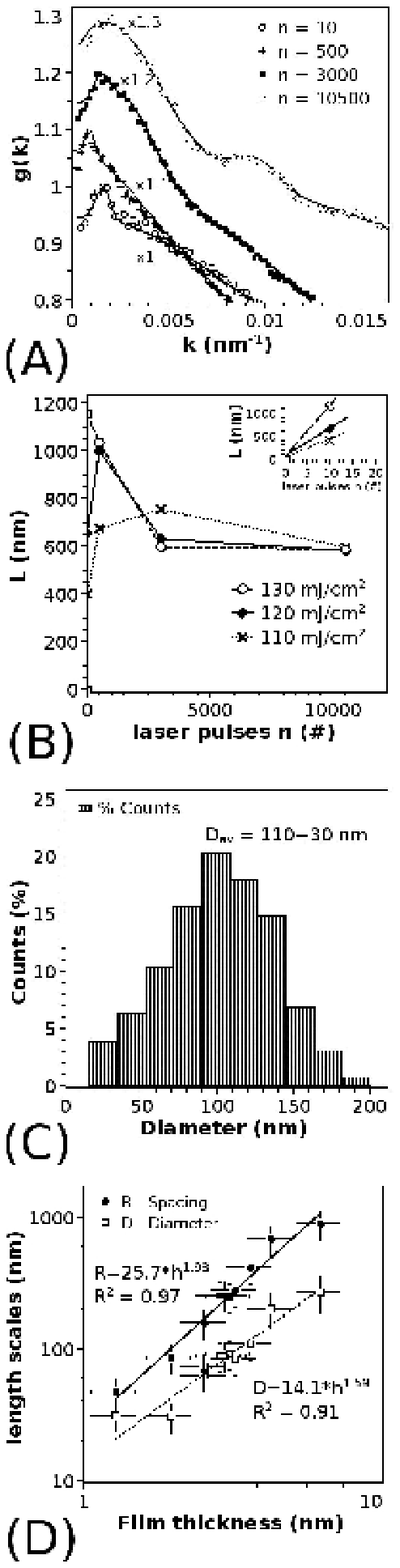}\par\end{centering}

\caption{\label{cap:Timedependence}}
\end{figure}

\end{document}